# Energetics of marine turbine arrays – extraction, dissipation and diminution*

Takafumi Nishino and Richard H. J. Willden

*Department of Engineering Science, University of Oxford, Parks Road, Oxford OX1 3PJ, UK*

## ABSTRACT

A two-scale modelling approach is discussed to predict the performance and energetics of a large number (more than a few hundred) of marine turbines installed as a power farm in a general coastal environment. The kernel of this approach is that the outer, or coastal-scale, model/simulation is to assess the reduction of flow passing through a given farm area as a function of the increase of head loss across the farm, whereas the inner, or device-scale, model/simulation employs this function to account for the (otherwise unknown) effect of coastal dynamics for that farm site, i.e. diminution of the power removed from the farm area due to the reduction of flow through the farm. Large-eddy simulations (LES) of periodic open channel flow (with a porous plate model representing turbines) are then presented as a device-scale part of such a two-scale model of large marine turbine arrays. Results demonstrate the usefulness of this approach to study how the overall energetics of turbine arrays (i.e. extraction, dissipation and diminution of energy within the entire farm area) may change depending on the characteristics of the farm site, array configuration and operating conditions of the turbines in the farm.

---

* This is an extended and revised version of a paper presented at the International Symposium on Marine and Offshore Renewable Energy, Tokyo, Japan, October 2013.

E-mail address: takafumi.nishino@eng.ox.ac.uk

## 1. Introduction

One of the key issues in the development of tidal- and ocean-current power generation systems is to understand properly their prospective performance characteristics, not only for each power generation device but also for device arrays or farms as a whole. This is important for assessing the maximum power extractable from the ocean as well as for optimising the configuration of the arrays or farms. Their performance is often described, especially from a viewpoint of device developers, in terms of the power extracted by the devices as useful power relative to either: (i) 'available' kinetic power (i.e. the kinetic power of upstream flow through the frontal-projected area of devices); or (ii) power removed from the flow (i.e. sum of that extracted as useful power and that dissipated into heat). When deploying a large number of devices as a farm, however, we need to consider that the flow through the entire farm area may reduce (due to very-large-scale flow deflection, for example) and therefore both (i) available kinetic power and (ii) power removed from the flow in that farm area may diminish substantially.

The importance of the reduction of flow through tidal arrays or farms has been discussed in recent theoretical studies, but from a few different viewpoints. Garrett and Cummins [1] considered an ideal power-extracting fence blocking an entire cross-section of a tidal channel (i.e. neglecting the dissipation of energy due to wake mixing) and derived an analytical model describing the relationship between the flow reduction and the power extracted (for a given head difference across the channel). Vennell [2-4] combined the same type of tidal channel dynamics model with a 'confined flow' version of the actuator disk theory [5-7], thereby examined the optimal channel-flow reduction to maximise the power of a number of devices arrayed regularly across the entire channel span (with gaps between each device; i.e. taking account of the dissipation of energy due to device-scale wake mixing). Meanwhile, Nishino and Willden [8, 9] considered an array of devices installed only across part of the channel span (so that the channel flow may bypass the entire array) and derived a 'two-scale' confined flow version of the actuator disk theory (i.e. considering the dissipation of energy due to both device- and array-scale wake mixing). Although in [8, 9] the channel inflow (rather than the head difference across the channel) was held constant to ignore the channel dynamics effects, again the reduction of flow through the array area played a key role in determining the conditions to maximise the power extracted.

These analytical models described above are all useful for understanding the basic mechanisms that govern the performance characteristics of tidal arrays/farms in an ideal tidal channel situation. When considering marine turbine arrays/farms in more general coastal environments, however, it seems very difficult to predict their performance analytically. It should be noted that, although the 'partial array' model of Nishino and Willden [8, 9] is conceptually applicable to general coastal areas that are exposed to the open sea (by considering zero or very low global blockage ratios), this model still needs to be combined with a kind of 'coastal dynamics' model that properly accounts for the effects of seabed friction and flow inertia across the entire area of interest. Since these effects depend significantly on site-specific conditions (bathymetric features etc.) and are very difficult to model analytically, we usually need to rely on numerical simulations of the shallow-water equations; see e.g. Adcock et al. [10]. Three-dimensional Navier-Stokes simulations are, at the present time, computationally very expensive and impractical for such geographic-scale flow simulations.

Regardless of whether we model the coastal dynamics analytically or numerically (to be combined with a model of power-extracting devices, which may also be either analytical or numerical), one important question to be answered is if — and if so, how — we should account for the interaction of (essentially three-dimensional) device-scale dynamics and (approximately two-dimensional) coastal dynamics. For example, a large number of devices densely deployed as a farm could change the vertical shear of the mean flow and thereby change the bed friction *coefficient* (as well as the bed friction itself) across the entire farm area, depending on the array configuration as well as operating conditions of devices within the farm [11]. Note that if the friction coefficient changes then the power generation characteristics of that farm area also change (as if the seabed in that farm area was changed physically by adding or removing rocks etc.). Could such interaction be important for the performance of turbine arrays? If so, how should we combine a three-dimensional device-scale flow model/simulation with a two-dimensional coastal dynamics model/simulation to account for the effects of such interaction?

In this study, we discuss a two-scale modelling approach to understand/predict the performance of a large number (more than a few hundred) of marine turbines deployed as a farm in a general coastal environment. The outline of the two-scale approach, which is a combined device-scale and coastal-scale simulation, is described first in Section 2, followed by an example of the device-scale part of the approach (LES of periodic open channel flow with simplified device models) in Section 3. By using the results of LES (together with a model function representing hypothetical coastal-scale simulation results), in Section 4 we demonstrate how the two-scale approach would work to understand/predict the performance and energetics of large turbine arrays as a farm (i.e., extraction, dissipation and diminution of energy within the farm area). Finally, conclusions are given in Section 5.

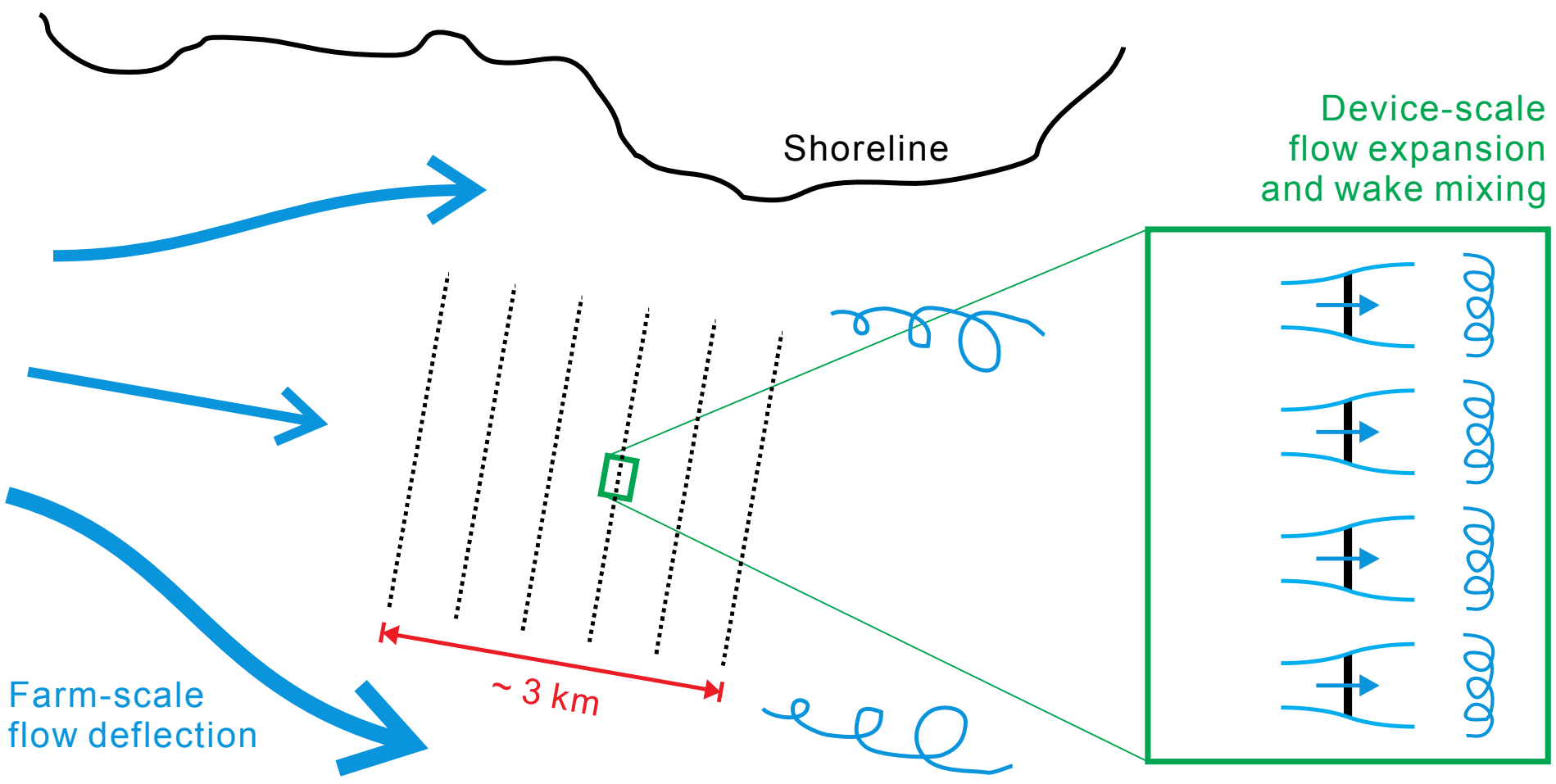

Fig. 1. Schematic of an example of coastal flow past a number (several hundred) of marine turbines.

## 2. Outline of the two-scale modelling of marine turbine arrays

Let us consider a large number (several hundred) of marine turbines regularly arrayed as a farm, as shown in Fig. 1. For the sake of convenience, we consider the size of each turbine to be about 20m (common size for megawatt class turbines) and that all turbines are deployed within a few kilometre square ‘farm area’. This seems a possible scenario, considering that the farm needs to co-exist with other coastal activities, such as shipping and fishing. The front and rear edges of the farm area are perpendicular to a quasi-steady ocean current.

In order to make the two-scale modelling feasible, we assume that the seabed is (at least approximately) flat across the entire farm area. In addition, we assume that all turbines in the farm operate under approximately the same local flow conditions; hence the ‘locally-averaged’ streamwise velocity (i.e. velocity averaged across each segment or portion of the farm in which each turbine is placed) is approximately the same as the mean velocity across the entire farm, $U_F$. We also assume that the head loss across the entire farm area, $H_F$, is much smaller than the average water depth in the farm area (say, less than a few percent of the average water depth, which is usually the case unless the Froude number of the flow is very high) and is (again at least approximately) uniform in the cross-stream direction. These assumptions need to be confirmed in coastal-scale simulations (based on the depth-averaged shallow water equations), where the effect of the farm may be modelled as an additional resistance (by increasing the bed friction coefficient, for example, although we might also need to consider the effect of additional turbulence generated by the turbines on the mixing of flow downstream of the farm area for better accuracy).

The purpose of the coastal-scale simulation here is *not* the prediction of the performance (power) of the farm but the assessment of coastal-dynamic characteristics of a given farm site. Specifically, we should obtain from coastal-scale simulations (by varying the flow resistance in the farm area) the *reduction* of the farm-averaged velocity, $\Delta U_F$, as a function of the *increase* in head loss across the farm, $\Delta H_F$. This information, $\Delta U_F = f(\Delta H_F)$, can be used for device-scale simulations to account for the effect of coastal dynamics (i.e. how much the current through the farm should decrease relative to the increase in the head loss) either rigorously as special inlet and outlet boundary conditions (interactively updated during each simulation) or only approximately as a scaling factor in the post-processing of the simulations (by assuming the Reynolds number independence of the flow). In this study we employ the latter approach, which will be demonstrated later in Section 4.

It should be noted that there are two extreme scenarios for the changes of $U_F$ and $H_F$ as the flow resistance in the farm area increases: (i) $U_F$ is fixed whereas $H_F$ increases; and (ii) $H_F$ is fixed whereas $U_F$ decreases. The first case is the 'fixed inflow' case, which is often assumed in studies focusing on device performance but is physically incorrect. The second case is the 'fixed head' case, which is again physically incorrect. Generally the changes of $U_F$ and $H_F$ must lie between the two cases, i.e. $U_F$ somewhat decreases whilst $H_F$ somewhat increases, depending on the coastal dynamics for that particular farm site. The smaller the decrease in $U_F$ (relative to the increase in $H_F$), the higher the power-generation potential in that area (for a given natural current speed). Hence the purpose of the coastal-scale simulation is, essentially, to select a high power-generation potential site (by assessing not only the natural current speed but also the relationship between $\Delta U_F$ and $\Delta H_F$), whereas the device-scale simulation is to examine how to make the most of the potential in that area.

## 3. LES of periodic open channel flow as a part of the two-scale modelling

In this section we present LES of periodic open channel flow (with a porous plate model representing turbines) as an example of the device-scale part of the two-scale modelling of marine turbine arrays. In this study we do not perform any coastal-scale simulations but assume that we have assessed the characteristics of some hypothetical farm sites, described by the form of $\Delta U_F = f(\Delta H_F)$. The effect of such farm-site characteristics on the performance and energetics of marine turbine arrays will be discussed later in Section 4.

### 3.1. *Hypothetical farm configuration*

In this study we consider a large farm consisting of multiple rows of turbines (often referred to as 'turbine fences') as illustrated earlier in Fig. 1. As an example, we consider turbines of the size of 20 metres arrayed regularly (with a lateral spacing of 20 metres between each turbine) within a three-kilometre-square farm area; each row or fence may therefore consist of up to 75 turbines. Here we presume that the streamwise spacing between each fence is large enough (relative to the size of each turbine) so that turbines in a downstream fence are not affected by the wake of individual turbines in upstream fences (the importance of this spacing will be discussed later). If we assume that this spacing needs to be at least 30 times larger than the turbine size, we may install up to six fences in this three-kilometre-square farm area.

It should be noted that this spacing between each fence (20m × 30 = 600m) is still much shorter than the fence width (3km). Hence we can expect that the local flow condition is nearly the same for most of the turbines in this farm area, which means that the two-scale modelling approach is approximately applicable to this type of farm. Strictly speaking, the flow through the farm area may somewhat decrease downstream (i.e. more flow may go through upstream fences than downstream fences) due to the leakage of flow through the spanwise ends of the farm area (unless the spacing between each fence is negligibly small compared to the fence width). A possible approach to account for the effect of such flow reduction through multiple fences has been reported in [12] but this is not considered in the present study.[1]

### 3.2. *Computational domain and conditions*

Figure 2 describes the computational domain for LES, which represents a (very small) part of the entire farm area. The depth, width and length of the domain are $H$, $4.8H$ and $12H$, respectively. Cartesian coordinates ($x$, $y$, $z$) are employed to represent the streamwise ($x$), vertical ($y$) and spanwise ($z$) directions. Each turbine is modelled as a porous plate, which is practically the same as the actuator disk model used earlier in [9] (except that the shape of the plate in this study is square rather than circular). The size of each plate, $d$, is 40% of the water depth and six plates are arrayed equidistantly (with spacing $s = d$) across the entire domain width. The 'hub-height' (the distance from the seabed to the centre of each turbine), $h$, is varied between 30% and 70% of the water depth $H$ in this study.

[1] Neglecting the non-uniformity of (locally-averaged) current speed within the farm area is expected to result in an underestimation of the farm output, essentially because the integral (over the farm area) of the cube of the actual (non-uniform) current speed is larger than that of the averaged (uniform) current speed $U_F$.

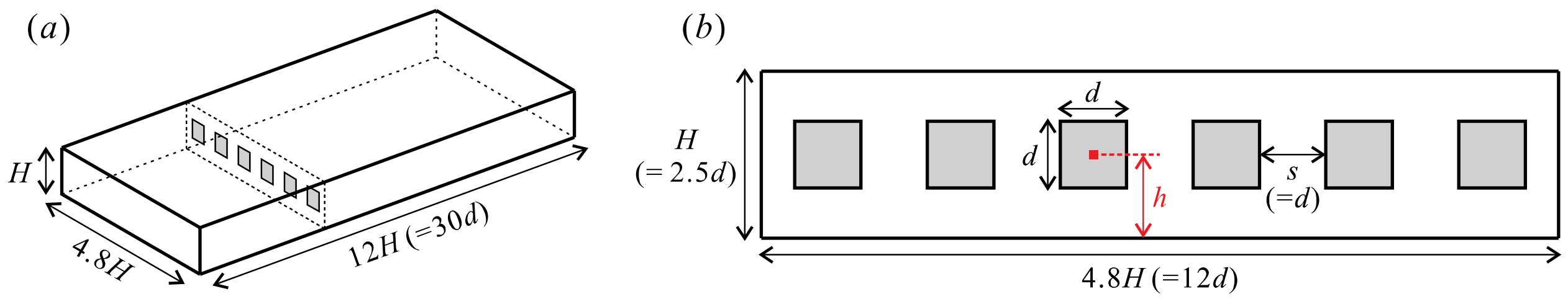

Fig. 2. Computational domain for LES of simplified marine turbines in a periodic open channel: (*a*) overview of the domain; (*b*) configuration of the array of turbines.

Here the blockage ratio $B = 0.2$ is the ratio of the total device area ($6d^2$) to the cross-sectional area of the domain ($30d^2$), whereas the device-to-bed area ratio $R = 0.01667$ is the ratio of the total device area to the bed area of the domain ($360d^2$). Note that, although not examined in this study due to the limitation of computational resources, both $B$ and $R$ are expected to be important for the performance of multiple rows of turbines.

One important aspect of the two-scale modelling in this study is that, for the device-scale simulation, we fix the mass flow through the domain and employ streamwise as well as spanwise periodic boundary conditions. We also employ the so-called 'rigid lid' boundary conditions for the sea surface, which means that we neglect the effect of local water height change within this small part of the farm on the performance of individual devices[2] (although we do obtain, as an important output of the simulation, the pressure difference between the inlet and outlet of this kinematically periodic domain). Therefore, for the device-scale simulation, the only independent flow parameter (apart from the configuration and resistance of the turbines) is the channel Reynolds number, $\mathrm{Re} = \rho U_{\mathrm{ref}} H/\mu$, where $\mu$ and $\rho$ are the viscosity and density of the water, respectively, and $U_{\mathrm{ref}}$ is the reference velocity (streamwise velocity averaged across the cross-section of the computational domain).

[2] Earlier studies have suggested that this is a reasonable assumption unless the local blockage ratio is extremely high [13].

In order to simulate directly a full-scale device flow ($d$ = 20m, $H$ = 50m and $U_{ref}$ = 2m/s, for example) we would need to perform LES of a very high Re (~ $10^8$) channel flow, which is computationally too expensive unless some forms of 'wall functions' are employed to model empirically the flow near the seabed. Instead of using such wall functions (which would not be very reliable when a strong streamwise pressure gradient is imposed by the turbines/plates), we perform LES at an only moderately high Re of 10935 (with no-slip conditions applied at the seabed) and assume that this Re is high enough to neglect the Reynolds number dependency of this device-scale flow. This assumption can be justified, at least partly, since at this Reynolds number our device models (porous plates) are still located entirely within the so-called 'outer layer' (i.e. $y^+ > 50$) where direct effects of viscosity on mean velocity profiles are negligible [14]. Also note that the device model used in this study is a simple porous plate (or actuator disk) model, which does not account for any Reynolds number effects on the device characteristics. Hence the results of the current LES at Re = 10935 (to be presented in Section 3.4) will be interpreted as an approximate representation of the device-scale flow at any practical channel Reynolds numbers (~ $10^8$) when discussing the performance and energetics of hypothetical farms later in Section 4.

### *3.3. Further details of the computations*

The LES computations were performed using ANSYS FLUENT (v14). The Smagorinsky eddy-viscosity model (with the model coefficient $C_s$ = 0.065, following the open channel LES study of Hinterberger et al. [15]) was used to model subgrid-scale stresses. Uniform structured grids with 300 and 240 cells were used in the $x$ and $z$ directions, resulting in resolutions of $\Delta x^+ \approx 24$ and $\Delta z^+ \approx 12$ in wall units, whereas in the $y$ direction 60 cells were used; 45 of which were uniformly allocated at $0.1 \leq y/H \leq 1$ and the other 15 were used to resolve the near-bottom region ($0 \leq y/H \leq 0.1$) with the minimum cell size of $\Delta y^+ \approx 1.2$ at the bottom. A second-order central difference scheme was used for spatial discretisation and a second-order implicit scheme was used for time integration. A constant non-dimensional time step of $U_{ref}\,\Delta t/H = 0.04$ was employed throughout the study. The computations were performed for 3000 time steps ($U_{ref}\,t/H$ = 120, i.e. 10 flow-through time for the periodic domain of 12$H$ length) first to obtain a statistically steady flow field, and then for another 3000 time steps to obtain statistics of the flow field.

The LES computations were performed first for the case without any turbines to compare the results with the open channel LES of Hinterberger et al. [15] at the same channel

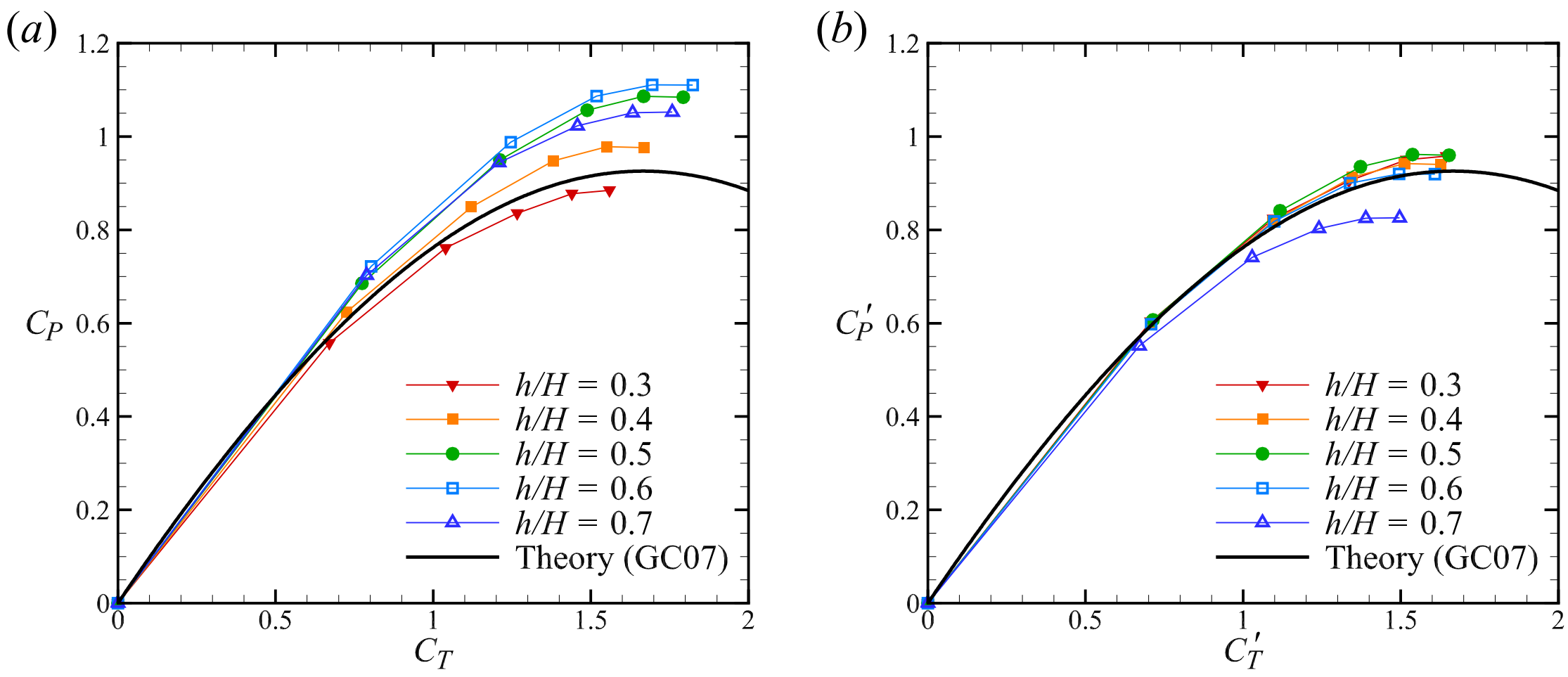

Fig. 3. Mean thrust and power coefficients: (*a*) original definitions; (*b*) modified definitions.

Reynolds number of 10935 for validation. The skin friction coefficient $C_{f0} = \tau_{w0}/(\frac{1}{2}\rho U_{\mathrm{ref}}^2)$ obtained was 0.00589, which agreed reasonably well with the value 0.00578 from [15] and also with the value 0.00582 from the direct numerical simulation of (closed rather than open) channel flow by Moser et al. [16]. Although not shown here for brevity, mean streamwise velocity and Reynolds stress profiles obtained in this validation case also agreed reasonably well with those from [15].

### *3.4. Results*

Figure 3(*a*) shows thrust and power coefficients of the devices (porous plates) obtained for five different hub-height cases ($h/H$ = 0.3, 0.4, 0.5, 0.6, 0.7). Plotted together is the solution of the analytical model of Garrett and Cummins [5] (hereafter GC07) for the same blockage $B = 0.2$. As with [9], the thrust and power coefficients, $C_T$ and $C_P$, are calculated as

$$C_T = \frac{[\text{device thrust}]}{\frac{1}{2}\rho U_{\mathrm{ref}}^2 \times [\text{device area}]} = K\frac{\langle U_d^2 \rangle}{U_{\mathrm{ref}}^2} \tag{1}$$

$$C_P = \frac{[\text{power extracted}]}{\frac{1}{2}\rho U_{\mathrm{ref}}^3 \times [\text{device area}]} = K\frac{\langle U_d^3 \rangle}{U_{\mathrm{ref}}^3} \tag{2}$$

where $K$ is the momentum loss factor (or 'resistance coefficient' of porous plates [7]); in this study we tested five different values ($K$ = 1, 2, 3, 4 and 5) for each hub-height case to change the thrust (note that $K$ = 2 is a theoretically optimum value yielding the well-known Betz limit of $C_P$ = 0.593 for an unconfined flow case, but for this confined flow case the optimum value is much larger than 2). $U_d$ is the time-averaged local streamwise velocity at the devices (porous plates), and $\langle U_d^2 \rangle$ and $\langle U_d^3 \rangle$ are the face averages of $U_d^2$ and $U_d^3$ over the device area, respectively.

As can be seen from Fig. 3(*a*), the maximum $C_P$ increases as $h/H$ increases from 0.3 to 0.6 but then somewhat decreases for $h/H = 0.7$. The increase in $C_P$ for $h/H = 0.3$ to 0.6 is due to the channel flow being faster at a higher position. This can be confirmed in Fig. 3(*b*), which plots modified thrust and power coefficients, $C'_T$ and $C'_P$, defined based on the square and cube of the undisturbed channel flow velocity averaged only across the turbine area for each hub-height case (rather than based on the square and cube of $U_{\mathrm{ref}}$ as in Eqs (1) and (2)). Note that the $C'_T$ versus $C'_P$ curves for $h/H = 0.3$, 0.4, 0.5 and 0.6 are nearly identical, and that they also agree well with the solution of the GC07 model.

Figure 3(*b*) also shows that the maximum $C'_P$ for $h/H = 0.7$ is about 10% lower than that for the other cases. The reason for this discrepancy is that, for this particular case, the mixing of device wake is rather slow and the streamwise gap between the turbine fences (30*d* in this study, due to the periodic boundary conditions) is not large enough for the wake to fully recover before reaching the turbines downstream. This can be seen from Figs 4 and 5, which show mean streamwise velocity profiles along the device centreline and across the water depth, respectively, for $h/H = 0.3$, 0.5 and 0.7. In Fig. 4 the results are shown for moderate ($K = 2$) and high ($K = 4$) resistance cases for each $h/H$, whereas in Fig. 5 the results are shown only for the moderate resistance case for each $h/H$. Note that $U_0$ used in Fig. 4 is the undisturbed mean streamwise velocity at each hub-height.

As noted earlier in the introduction, one potentially important issue for the design of marine turbine arrays is how the array configuration affects the bed friction coefficient for the entire farm area. Here we define the (average) bed friction coefficient as

$$\langle C_f \rangle = \frac{\langle \tau_{\mathrm{w}} \rangle}{\frac{1}{2}\rho U_{\mathrm{ref}}^2} \qquad (3)$$

where $\langle \tau_w \rangle$ is the wall shear stress averaged over the bottom wall of the entire computational domain. If the bed friction coefficient is higher for a certain array configuration, then more power tends to be lost/dissipated for this array configuration (as if some rocks, obstacles or additional device-supporting structures were brought into the farm area). Figure 6(*a*) shows the obtained friction coefficient $\langle C_f \rangle$ normalised by that for the case with no turbine thrust, $C_{f0}$ = 0.00589, for the five different *h/H* cases. Note that, although the Reynolds number for the present LES is about four orders-of-magnitude smaller than that for a full-scale device flow, this $C_{f0}$ value is still in the same order-of-magnitude as that for a full-scale device flow with some roughness on the seabed (typically between 0.004 and 0.02). As can be seen from the figure, the friction coefficient was found to increase by up to about 30 ~ 40% as the turbine thrust was increased, and the rate of increase was found to depend (albeit moderately) on the hub-height of the turbines (highest for *h/H* = 0.7 and lowest for *h/H* = 0.5).

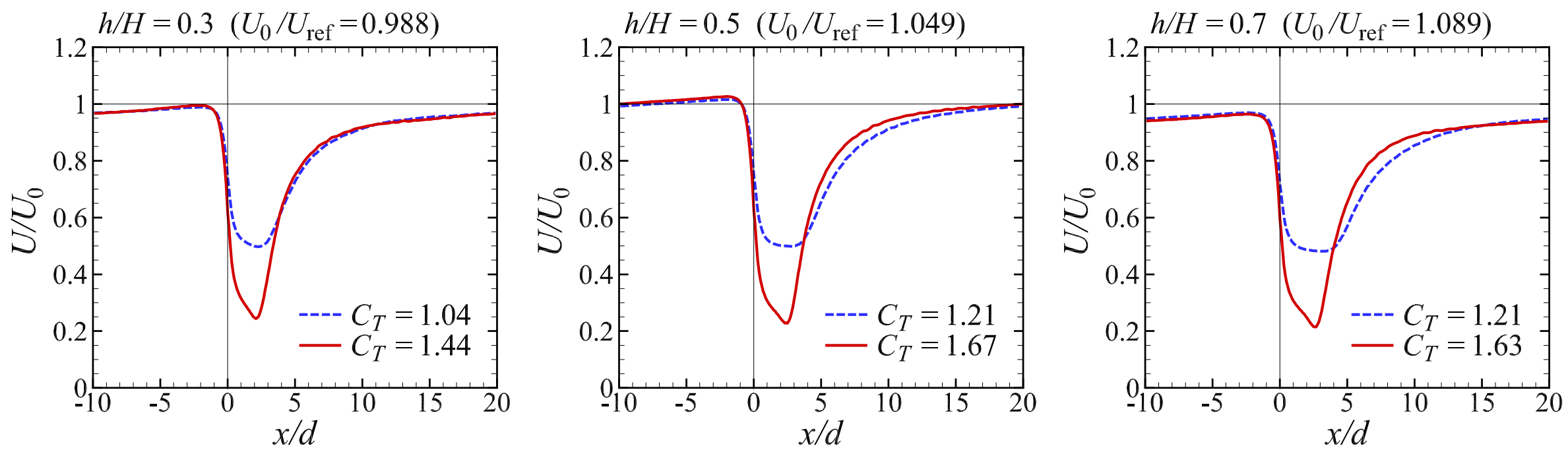

Fig. 4. Mean streamwise velocity along the device centreline: for *h/H* = 0.3, 0.5 and 0.7. Dashed and solid lines show the results of moderate ($K$ = 2) and high ($K$ = 4) resistance cases, respectively.

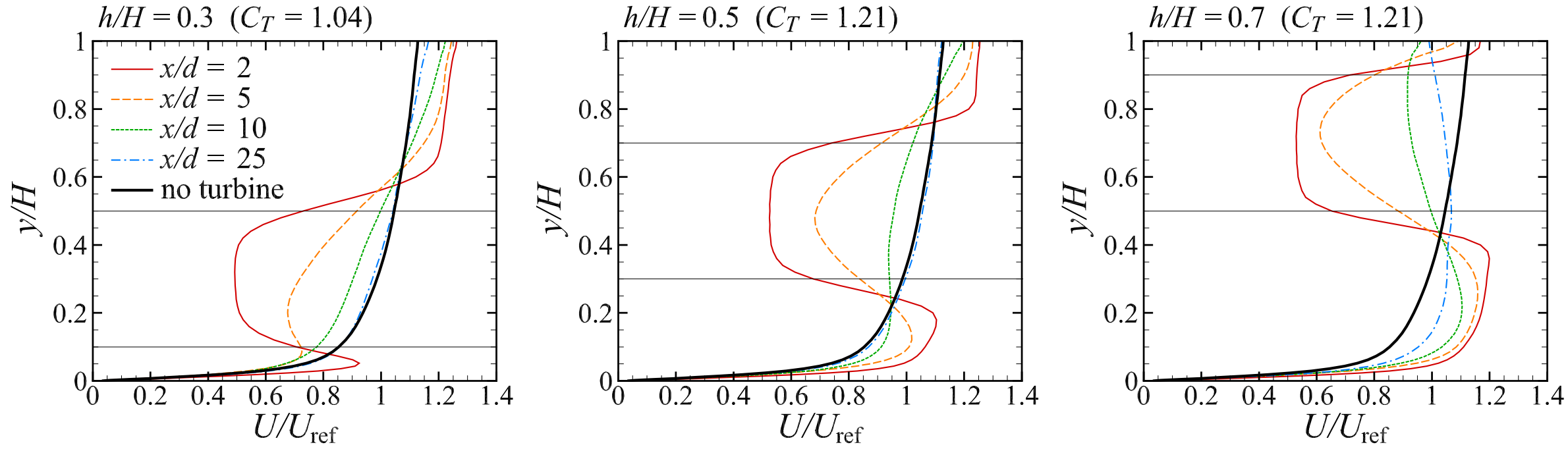

Fig. 5. Mean streamwise velocity profiles across the water depth downstream of the centre of each device (at *x/d* = 2, 5, 10 and 25) for moderate ($K$ = 2) resistance cases: for *h/H* = 0.3, 0.5 and 0.7.

It should be noted, however, that the increase in the bed friction coefficient is not spatially uniform. Figure 7 shows contours of instantaneous and time-averaged values of the friction coefficients for *h/H* = 0.3, 0.5 and 0.7 (with $K$ = 2). For *h/H* = 0.3, the friction coefficient is locally very high immediately downstream of the device location ($x/d$ = 0). As *h/H* is increased to 0.5 and 0.7, this local peak value gradually decreases. For *h/H* = 0.7, however, the friction level is high in the far downstream region ($x/d$ > 15 and $x/d$ < 0) compared to the other two cases. These differences in the spatial variation of $C_f$ suggest that the order of different hub-height cases for the rate of increase in $\langle C_f \rangle$, shown in Fig. 6(*a*), might change depending on the streamwise spacing between the turbine fences (or on the device-to-bed area ratio $R$). It should also be noted that the porous plate model used in this study is very simple and might not have simulated accurately the effect of turbine arrays on the bed friction coefficient. It would therefore be informative to perform a further analysis on the bed friction coefficient using a more sophisticated turbine model, such as an actuator line model (e.g. [17]), in future studies.

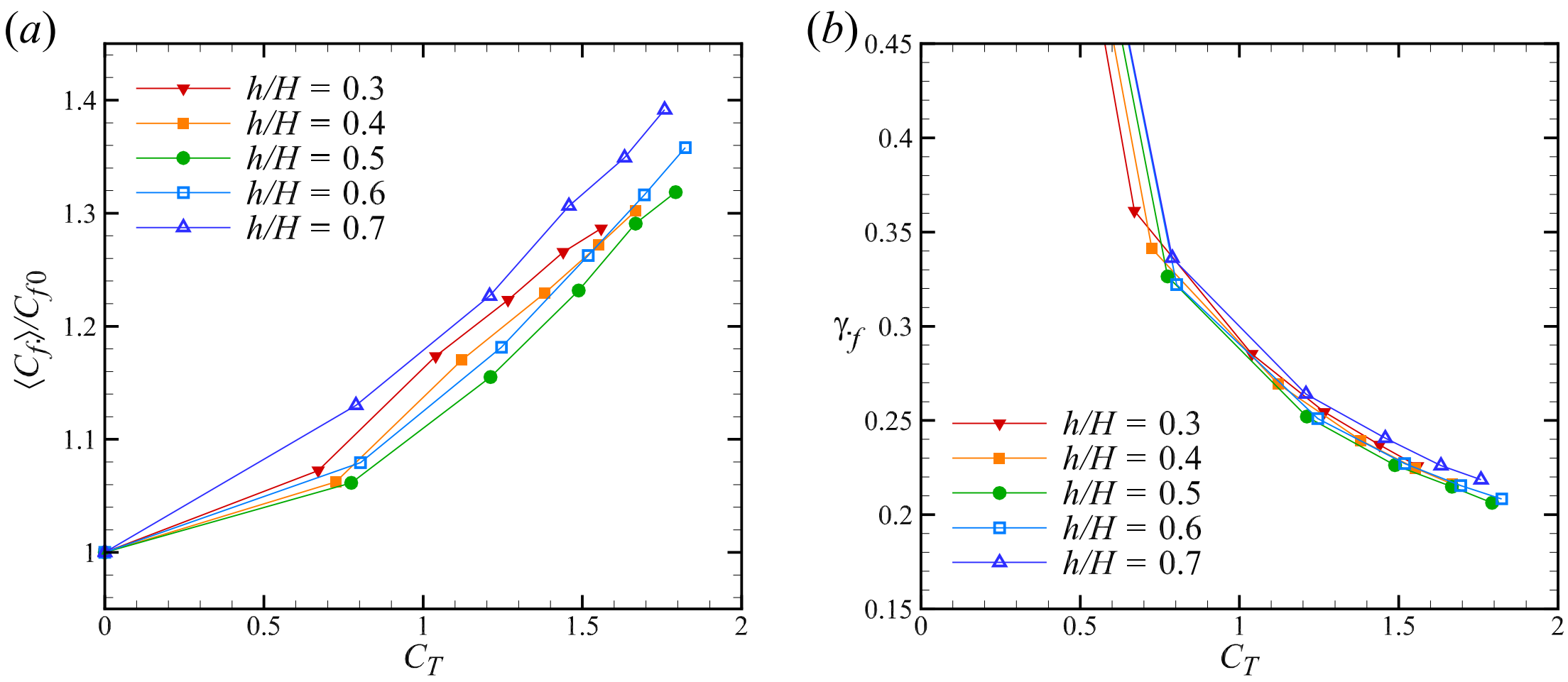

Fig. 6. Effects of turbine thrust on: (*a*) normalised bed friction coefficient averaged across the entire channel, and (*b*) the ratio of the bed friction drag to the total drag.

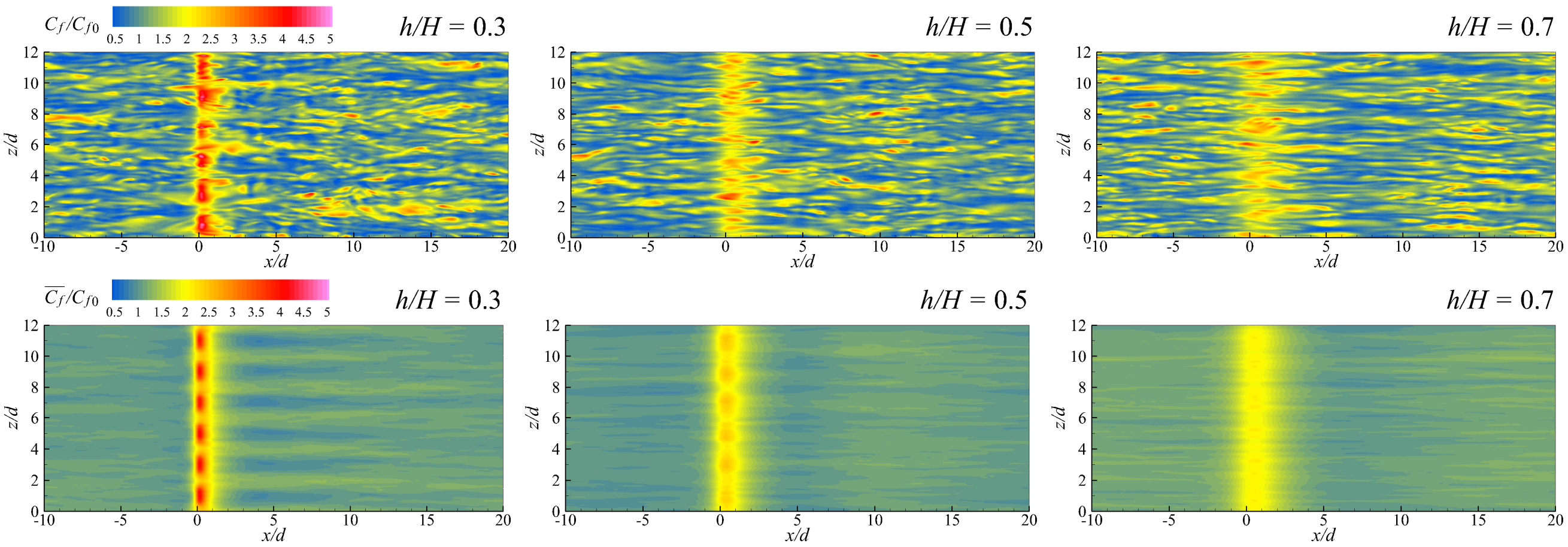

Fig. 7. Contours of skin friction on the bottom of the channel: (top) instantaneous and (bottom) time-averaged results for moderate ($K = 2$) resistance cases, for $h/H = 0.3$, 0.5 and 0.7.

Also of importance is that, although the bed friction coefficient $\langle C_f \rangle$ increases with the turbines' thrust coefficient $C_T$ as shown earlier in Fig. 6(*a*), this increase in $\langle C_f \rangle$ is (at least in the present study using the porous plate model) rather gentle relative to the increase of the turbine thrust. Figure 6(*b*) presents the ratio of the bed friction drag to the total drag, $\gamma_f$, which is calculated by

$$\gamma_f = \frac{[\text{bed friction drag}]}{[\text{total drag}]} = \frac{\langle C_f \rangle}{\langle C_f \rangle + RC_T} \tag{4}$$

where $R$ (= 0.01667 in this study) is the device-to-bed area ratio defined earlier in Section 3.2. It can be seen that this ratio decreases down to about 0.2 as $C_T$ increases, i.e. the bed friction drag accounts for only about 20% of the total drag for high thrust cases simulated in this study. This essentially means that, as will be shown later in Section 4, the power loss due to seabed friction becomes less important (compared to that due to device wake mixing) as the turbine thrust increases.

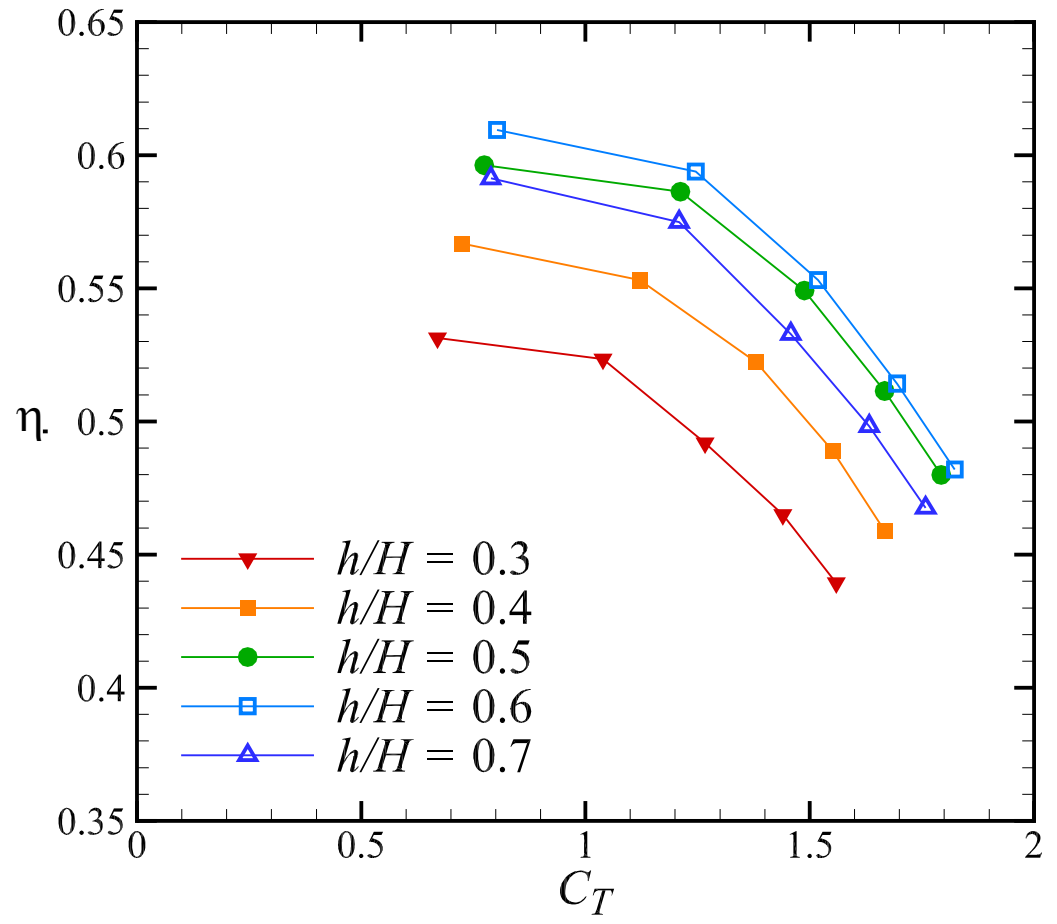

Fig. 8. Effect of turbine thrust on the energy conversion efficiency of the device array.

Finally, Fig. 8 shows the effect of turbine thrust on the energy conversion efficiency (sometimes referred to as 'basin efficiency') of the device array:

$$\eta = \frac{[\text{power extracted}]}{[\text{power removed from the flow}]} = \frac{RC_P}{\langle C_f \rangle + RC_T} \ . \tag{5}$$

As can be seen from the above equation, the efficiency is zero when $C_T$ is zero (since $C_P$ is zero whilst $\langle C_f \rangle$ is non-zero). However the efficiency increases very rapidly (up to about 0.6, depending on $h/H$) as $C_T$ and $C_P$ increase from zero to non-zero values, and then decreases gradually as $C_T$ further increases. For the sake of convenience, here we can also define another type of power coefficient representing the power removed from the flow (i.e. sum of that extracted as useful power and that dissipated into heat):

$$C_{P+} = \frac{[\text{power removed from the flow}]}{\frac{1}{2}\rho U_{\text{ref}}^{3} \times [\text{device area}]} = \frac{C_P}{\eta} \ . \tag{6}$$

As will be shown later in Section 4, the energy conversion efficiency $\eta$, or alternatively $C_{P+}$, is important for understanding the performance and energetics of large marine turbine arrays that cause the reduction of flow through the entire farm area.

## 4. Energetics of marine turbine arrays

In the following we demonstrate how the results of device-scale simulations can be used to assess the performance and energetics of multiple rows of devices deployed at a given farm site. As described earlier, the basic idea underlying the two-scale coupling in this study is that we assume the Reynolds number independence of device-scale flow dynamics, so that the LES results presented earlier (namely $C_P$, $C_T$ and $\langle C_f \rangle$ obtained for various $h/H$ and $K$) represent those at any practical Reynolds numbers ($10^7 \sim 10^8$). Specifically, here we assume that the same LES results ($C_P$, $C_T$ and $\langle C_f \rangle$ for given $h/H$ and $K$) are obtained for the following dimensional parameters: $d$ = 20m, $H$ = 50m and $U_{\text{ref}} = U_F = 1 \sim 2$m/s. Note that a natural interpretation of $H$ is the average water depth for the entire farm area, although this may alternatively be interpreted as a local water depth in any part of the farm, provided that the total head loss across the farm ($H_F$) is always much smaller than this water depth (even for high turbine-thrust cases). Meanwhile, the mean current velocity across the farm ($U_F$) may reduce significantly from that for the zero turbine-thrust case, $U_{F0}$ (= 2m/s).

As already mentioned briefly in Section 2, we define $\Delta U_F$ as the reduction of $U_F$ and $\Delta H_F$ as the increase of $H_F$ (compared to the zero turbine-thrust case); hence the following relationships exist:

$$U_F = \left(U_{F0} - \Delta U_F\right) \tag{7}$$

$$H_F = \left(H_{F0} + \Delta H_F\right) << H \tag{8}$$

where $H_{F0}$ is the value of head loss $H_F$ for the zero turbine-thrust case. Here we consider deriving the relationship between $U_F$ and $H_F$ from the results of device-scale simulations first. Although the effect of local water height change was neglected in our LES employing the 'rigid lid' assumption, we can still estimate the total head loss across the farm ($H_F$) from the pressure loss between the inlet and outlet of the LES computational domain, $P_{\text{in}} - P_{\text{out}}$, multiplied by the number of turbine rows in the farm, $n_{\text{row}}$, as

$$H_F = n_{\text{row}} \left( \frac{P_{\text{in}} - P_{\text{out}}}{\rho g} \right) \tag{9}$$

where $g$ = 9.8m/s$^2$ is the acceleration due to gravity. Here the pressure loss can be calculated for a given $U_F$ (by considering the momentum conservation within the LES domain and also the Reynolds number independence of $\langle C_f \rangle$ and $C_T$) as

$$P_{\text{in}} - P_{\text{out}} = \frac{1}{2} \rho U_F^2 \frac{B}{R} \left( \langle C_f \rangle + R C_T \right) \tag{10}$$

and hence, from Eqs (9) and (10), we obtain

$$H_F = \frac{n_{\text{row}}}{2} \frac{U_F^2}{g} \frac{B}{R} \left( \langle C_f \rangle + R C_T \right) \tag{11}$$

or alternatively

$$\frac{H_F}{H} = \frac{n_{\text{row}}}{2} \left( \text{Fr}_0 \right)^2 \frac{B}{R} \left( \langle C_f \rangle + R C_T \right) \left( \frac{U_F}{U_{F0}} \right)^2 \tag{12}$$

where $\text{Fr}_0 = U_{F0} / \sqrt{gH}$ = 0.0904 is the Froude number of the flow for the case with no turbine thrust. As an example, Fig. 9(*a*) shows solutions of Eq. (12) obtained for the medium hub-height case $h/H = 0.5$ (with five different device conditions, $K = 1, 2, 3, 4, 5$) and $n_{\mathrm{row}} = 6$. Note that we can also calculate $H_{F0}$ from Eq. (12) by considering $\langle C_f \rangle = C_{f0}$, $C_T = 0$ and $U_F/U_{F0} = 1$.

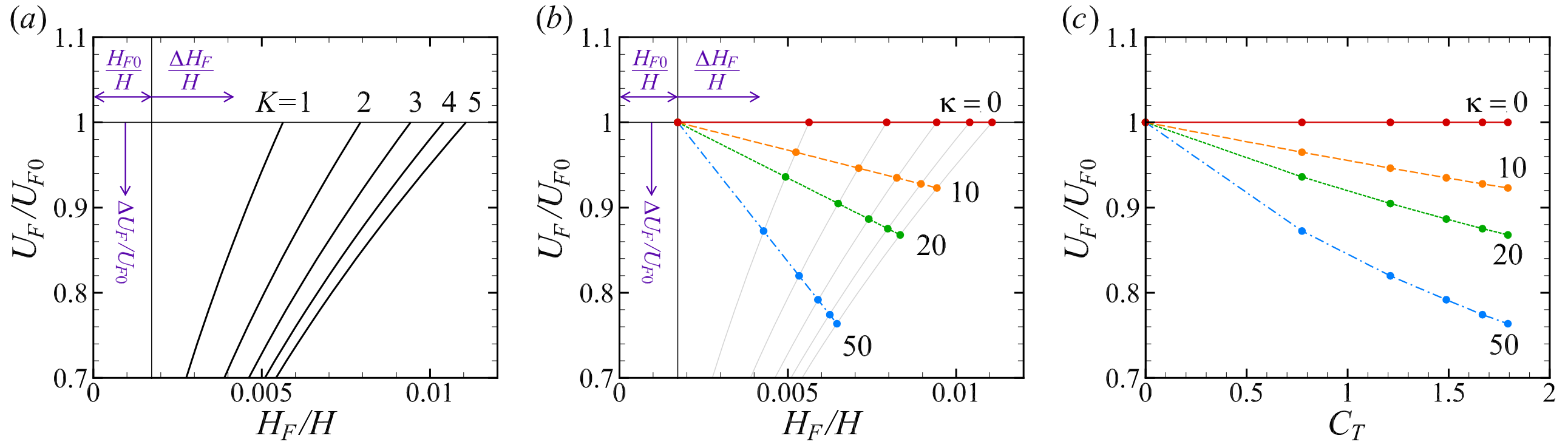

Fig. 9. Two-scale coupling process to determine the farm flow reduction for four hypothetical farm sites ($\kappa$ = 0, 10, 20 and 50), for $h/H$ = 0.5 and $n_{\text{row}}$ = 6: (*a*) $U_F$ vs. $H_F$ obtained from the device-scale problem (i.e. solutions of Eq. (12)) for $K$ = 1, 2, 3, 4 and 5; (*b*) $U_F$ vs. $H_F$ obtained from the coastal-scale problem (i.e. solutions of Eq. (14)); and (*c*) determined farm flow velocities plotted against $C_T$.

Meanwhile, the relationship between $U_F$ and $H_F$ can be obtained from coastal-scale simulations as well, as mentioned earlier in Section 2. For demonstration purposes, here we consider that the relationship between $\Delta U_F$ and $\Delta H_F$ can be approximated (when $\Delta U_F$ is not very large, say $\Delta U_F/U_{F0} < 0.3$) by the following linear function:

$$\frac{\Delta U_F}{U_{F0}} = \kappa \frac{\Delta H_F}{H} \tag{13}$$

where $\kappa$ is a site-specific parameter describing how much the flow through the entire farm tends to reduce relative to the increase in the head loss across the farm (note that $\kappa = 0$ and $\infty$ correspond to the ideal 'fixed inflow' and 'fixed head' scenarios, respectively). Hence, by substituting Eqs (7) and (8), we obtain

$$\frac{U_F}{U_{F0}} = 1 - \kappa\left(\frac{H_F}{H} - \frac{H_{F0}}{H}\right). \tag{14}$$

Figure 9(*b*) shows solutions of Eq. (14) for four different hypothetical farm sites with $\kappa = 0$, 10, 20 and 50. Plotted together in this figure (with light gray lines) are the solutions of Eq. (12) shown earlier in Fig. 9(*a*). Since both Eqs (12) and (14) must be satisfied, these are the two equations to be solved simultaneously to determine $U_F$ and $H_F$ (as indicated by symbols in Fig. 9(*b*)) for a given set of device-scale and coastal-scale conditions (array configuration, operating conditions of devices and farm-site characteristics). The determined $U_F$ values (for the four hypothetical farm sites) are re-plotted against the devices' thrust coefficient in Fig. 9(*c*), showing the importance of farm-site characteristics (represented by $\kappa$) on the reduction of flow through the entire farm area.

As the farm flow reduction has been determined successfully by coupling the device-scale and coastal-scale problems, we can now analyse the overall energetics (energy budgets) of turbine arrays as a farm. For the sake of convenience, here we introduce the farm flow reduction factor:

$$\alpha = \frac{U_F}{U_{F0}} \tag{15}$$

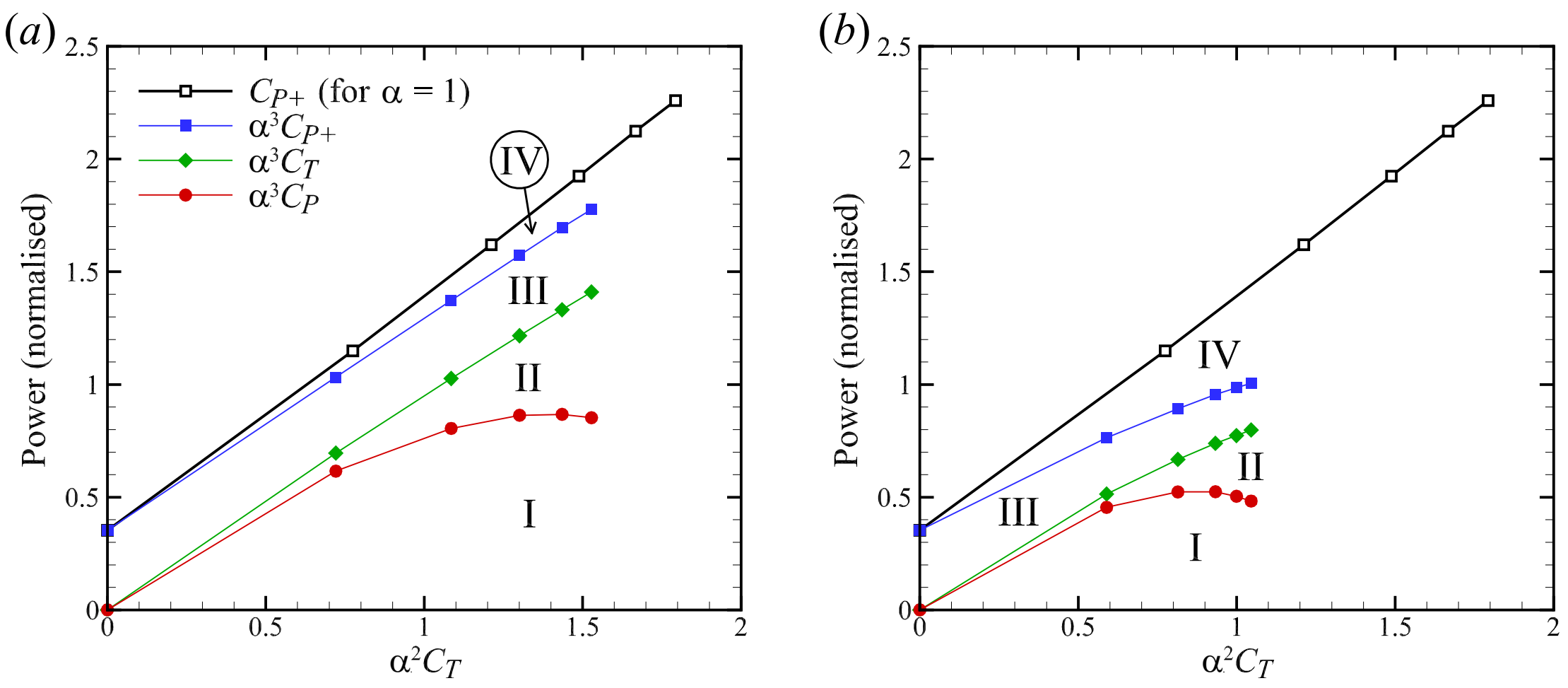

Fig. 10. Energetics of marine turbine arrays installed in: (*a*) high potential ($\kappa = 10$) and (*b*) low potential ($\kappa = 50$) farm sites, for $h/H = 0.5$ and $n_{\mathrm{row}} = 6$. The meanings of Regions I to IV are as follows. I: power extracted as useful power, II: power dissipated due to device wake mixing, III: power dissipated due to bed-induced shear, and IV: diminution of the total power removed from the farm area due to the reduction of flow through the farm.

so that the 'global' thrust and power coefficients of devices deployed in the farm (i.e. the coefficients defined using the undisturbed current speed $U_{F0}$ rather than $U_F$) are $\alpha^2 C_T$ and $\alpha^3 C_P$, respectively, in analogy with those defined in [8, 9]. Also, the 'global' version of $C_{P+}$ defined earlier in Eq. (6) is now described as $\alpha^3 C_{P+}$. Note that both $\alpha^3 C_P$ and $\alpha^3 C_{P+}$ may represent power (extracted or removed) for either one device or all devices, relative to the kinetic power of undisturbed flow through the *corresponding* device area (i.e. area for one device or all devices). In addition, since $C_{P+} = C_T$ when the bed friction loss is zero (note that $\eta = C_P/C_T$ when $\langle C_f \rangle = 0$), we can consider $\alpha^3 C_T$ to represent the sum of the power extracted as useful power and that lost due to device wake mixing (again relative to the kinetic power of undisturbed flow through the corresponding device area).

Two examples of the energetics of turbine arrays, deployed as a farm in high ($\kappa = 10$) and low ($\kappa = 50$) potential sites, are presented in Fig. 10 (for $h/H = 0.5$, $n_{\mathrm{row}} = 6$). Note that the vertical axis shows power (for either one device or all devices) normalised by the kinetic power of undisturbed flow through the corresponding device area, whilst the horizontal axis shows the global thrust coefficient of devices. Here we can consider four different factors contributing to the energetics of device arrays, namely Regions I to IV denoted in Fig. 10: Region I represents the power extracted as useful power, Region II is the power dissipated due to device wake mixing, Region III is the power dissipated due to bed-induced shear, and Region IV is the diminution of the total power removed from the farm area due to the farm flow reduction (compared to the case where the flow through the farm does not reduce at all, i.e. $\alpha = 1$, for any device thrust conditions). As can be seen from the figure, the diminution of the power removed from the farm (Region IV) is small when the devices are deployed in the high potential farm site ($\kappa = 10$), but this increases significantly in the low potential farm site ($\kappa = 50$). Of importance here is that, as $\kappa$ increases and thus Region IV increases, Regions I to III diminish all together, i.e. the power extracted as useful power and that dissipated into heat both decrease, since they are all proportional to $\alpha^3$. As a result, the optimal operating condition (or thrust condition) of devices to maximise the output power (Region I) changes significantly depending on the value of $\kappa$ or the farm-site characteristics.

Finally, it should be remembered that, since we employed an ideal porous plate (or actuator disk) model to simulate the effect of devices, the 'extracted' power (Region I) here represents the *limit* of power extraction. For actual marine turbines, some of this 'extracted' power (Region I) would not be extracted but be wasted, either (i) by generating heat at the turbines due to mechanical losses etc., or (ii) by generating additional fluid motions, such as mean swirling, large-scale coherent motions and turbulence (sometimes referred to as blade-induced turbulence or BIT [18]), the power of which is also eventually dissipated into heat due to additional wake mixing [19]. Such additional loss may also be caused due to device-supporting structures. It would therefore be useful to perform a similar two-scale analysis using a more practical turbine model in future studies. Also note that the diminution of the power removed from the farm (Region IV) can be due to two different types of reduction of flow through the farm, depending on how large the drag caused by the farm is compared to the driving force of the ocean current. If the farm drag is negligibly small compared to the driving force (which is usually the case), the farm flow reduction is essentially due to very-large-scale flow deflection (as illustrated in Fig. 1), which generally results in additional mixing and thus an increase in dissipation somewhere outside of the farm. However, if the farm is extremely large and the farm drag is not negligibly small compared to the driving force, the farm would somewhat decelerate the ocean current and therefore the flow through the farm may reduce even without any flow deflection around the farm.

## 5. Conclusions

In this study we discussed a generic two-scale modelling approach to understand/predict the performance and energetics of a large number (more than a few hundred) of marine turbines forming a hydro-power farm in a general coastal environment. Similarly to the theoretical partial tidal fence model recently proposed in [8], this approach consists of two individual parts (simulations) of different scales, namely the device-scale and coastal-scale simulations. The purpose of the coastal-scale simulation is not the prediction of the performance (output power) of the farm but to assess the reduction of flow through the farm as a function of the increase of head loss across the farm. This information is used to account for the so-called coastal dynamics effects (for that particular farm site) when analysing the results of device-scale simulations (simulating only a small part of the entire farm) to eventually predict the performance and energetics of large turbine arrays deployed as a farm.

To demonstrate how the present two-scale modelling approach would work, we have also performed LES of periodic open channel flow over a spanwise array of six devices (represented by porous plate models, with various hub-height) as an example of device-scale simulations, whilst employing a simple model function to account for the coastal dynamics effects for four hypothetical farm sites. Results demonstrated that this approach can be used to study effectively how the energetics of a large number of devices as a farm (i.e. extraction, dissipation and diminution of energy within the entire farm area) may change depending on the farm site characteristics, array configuration (hub-height in this example) and device operating conditions, although the influence of the hub-height was found to be relatively minor in this example. Most importantly, this approach would be very useful to determine the optimal operating (or thrust) condition of a large number of devices to maximise their total output power for a given farm site and for a given array configuration.

Finally, it should be noted that this two-scale modelling/analysis is easily applicable to many other device-scale simulations using more practical device models, such as actuator line models and even fully-resolved rotor models with device-supporting structures. Such device-scale simulations coupled with the results of coastal-scale simulations would provide a more practical and reliable prediction of the performance and energetics of marine turbine arrays deployed as a large farm.

**Acknowledgement**

The authors gratefully acknowledge the support of the Oxford Martin School, University of Oxford, who have funded this research.